\pdfoutput=1
\documentclass{sig-alternate}
\makeatletter
\newif\if@restonecol
\makeatother

\usepackage{graphicx}
\usepackage{amsmath}
\usepackage{amssymb}
\usepackage[boxed,lined]{algorithm2e}
\begin{document}
%
\conferenceinfo{CIKM'08,} {October 26--30, 2008, Napa Valley, California, USA.} 
\CopyrightYear{2008}
\crdata{978-1-59593-991-3/08/10} 

\title{An Algorithm to Determine Peer-Reviewers}

\numberofauthors{2} 
\author{
\alignauthor
Marko A. Rodriguez\\
       \affaddr{Digital Library Research and Prototyping Team}\\
       \affaddr{Los Alamos National Laboratory}\\
       \affaddr{Los Alamos, New Mexico 87545}\\
       \email{marko@lanl.gov}
\alignauthor
Johan Bollen\\
       \affaddr{Digital Library Research and Prototyping Team}\\
       \affaddr{Los Alamos National Laboratory}\\
       \affaddr{Los Alamos, New Mexico 87545}\\
       \email{jbollen@lanl.gov}
}
\date{30 May 2008}

\maketitle
\begin{abstract}
The peer-review process is the most widely accepted certification mechanism for officially accepting the written results of researchers within the scientific community.  An essential component of peer-review is the identification of competent referees to review a submitted manuscript. This article presents an algorithm to automatically determine the most appropriate reviewers for a manuscript by way of a co-authorship network data structure and a relative-rank particle-swarm algorithm. This approach is novel in that it is not limited to a pre-selected set of referees, is computationally efficient, requires no human-intervention, and, in some instances, can automatically identify conflict of interest situations. A useful application of this algorithm would be to open commentary peer-review systems because it provides a weighting for each referee with respects to their expertise in the domain of a manuscript. The algorithm is validated using referee bid data from the 2005 Joint Conference on Digital Libraries.
\end{abstract}

\section{Introduction}

The peer-review process is the {\it de facto} standard for validating the written results of researchers within the scientific community. In its present form, the peer-review process is mediated by journal editors and/or conference organizers. They receive manuscripts from authors, identify competent referees to review the manuscripts, and ultimately accept or reject each manuscript for publication or presentation on the basis of referee feedback. In the chain leading from a manuscript's submission to an editor's decision, the identification of competent referees constitutes a crucial first step; it will shape the quality and reliability of the subsequent reviewing.

Referee identification has mainly been a human-driven process; editors and conference organizers rely on their subjective assessments of a particular domain and the submission's content to identify a set of appropriate referees. However, it is not at all certain that editors have complete knowledge of all potentially competent referees for a particular manuscript, and, even if that were the case, that they are always able to produce an objective, good match between the manuscript and this pool of potential referees. Research has in fact indicated the peer-review process is subject to numerous sources of biases and unreliability, many of which are undoubtedly caused by mismatches between a manuscript and its referees \cite{mapping:rodriguez2007}. Furthermore, with the advent of open commentary peer-review systems for pre-print repositories \cite{peer:rodriguez2005} such as Naboj\footnote{Naboj available at: http://www.naboj.com/} and web journals such as Interjournal\footnote{InterJournal available at: http://www.interjournal.org/} and Philica\footnote{Philica available at: http://www.philica.com/}, the requirements for an efficient peer-review process has changed. When any reader can submit a review, separating the `wheat from the chaff' becomes a high priority to validly assess the quality of a manuscript. Locating referees to review a specific manuscript is thus gradually becoming less important as identifying which of the many provided reviews originate from actual experts in the manuscript's domain.

A number of automated referee identification algorithms have been proposed in the literature to more objectively and efficiently match a submitted manuscript to a set of competent, i.e., expert referees. Previously published algorithms have mostly relied on matching referee-provided textual indicators of interest, e.g. key terms,  to the contents of manuscripts. Dumais et al (1992) and Yarowsky et al (1999) \cite{dumais:peer1992,yarowsky:peer1999} use Latent Semantic Indexing (LSI) to match manuscript abstract to referees. Other approaches determine referee expertise via web mining techniques \cite{basu:peer2001}, and/or asking authors and the referees to provide keyterms describing their manuscript and area of expertise respectively \cite{guervo:peer2004}. However, it is not feasible to require all individuals in the scientific community to report on their interest and expertise in this manner. Nor is it feasible to perform latent semantic indexing on the websites and/or articles of all scientists in the community due to costs associated with text analysis on a large data set. Applications of the mentioned referee identification algorithms have therefore been restricted to situations in which such information can be obtained for a pre-selected set of individuals, e.g. conferences and workshops. They have consequently failed to gain acceptance in the domain of classic journal peer-review and open commentary peer-review.

This article proposes a referee identification algorithm that is both computationally inexpensive and requires no intervention on behalf of the authors, journal editors, and/or conference organizers. The proposed algorithm identifies appropriate referees for a manuscript by applying a particle-swarm algorithm to a co-authorship network. A particle-swarm is a discrete form of the spreading activation algorithm of information retrieval \cite{spread:collins1975,applic:crestani1997}. In short, the proposed algorithm provides a context-specific weight for every individual represented in the co-authorship network, where the context is the paper required for review. The context-specific aspect of the algorithm places the algorithm into the class of relative rank algorithms (i.e.~ranking with priors) \cite{markov:white2003}.  Furthermore, this context-sensitive weighting provides a strong incentive for its use in open commentary peer-review. To date, no such referee weighting algorithm has been proposed in the literature.

The algorithm's performance is validated against referee bid data provided by the program chair and steering committee of the 2005 Joint Conference on Digital Libraries (JCDL) \cite{sumner:jcdl2005}. We show how the algorithm can properly identify appropriate referees and, in some cases, conflicts of interests, and suggest how its accuracy can be improved by including additional data sources. 

\section{The Proposed Referee Identification Algorithm}

The referee identification algorithm presented in this paper is dependent upon: 
\begin{enumerate}
	\item a co-authorship network data structure
	\item a relative-rank particle-swarm propagation algorithm
\end{enumerate}

Our approach is based on the premise that a manuscript's subject domain can be represented by the authors of its references. Starting from those authors, we can identify related authors in a co-authorship network who may be potential referees for the submitted manuscript. To locate such related authors, a particle-swarm starting, from the referenced authors, diffuses an energy distribution over a co-authorship network in a manner similar to the spreading activation techniques used for information retrieval \cite{applic:crestani1997}, but in a discrete form related to the random walker algorithms of Markov chain analysis \cite{inside:bianchini2005}. However, unlike the iterative algorithms that identify a stationary distribution such as PageRank \cite{google:brin1998} and eigenvector centrality \cite{power:bonacich1987,socialanal:wasserman1994}, the proposed algorithm does not generate nor presuppose a particular network topology (e.g.~aperiodic and connected). PageRank and eigenvector centrality algorithms are global rank metrics in that the initial distribution of energy in the network does not effect the final energy distribution when the algorithm has converged to a steady state vector. Instead, the proposed algorithm is a relative rank algorithm in that the initial distribution of energy, or particles, in the network determines the final author ranking \cite{markov:white2003}. The relative rank algorithm proposed in \cite{markov:white2003} uses a ``back probability" to allow walkers to ``teleport" to their original source node. In this manner, a steady state vector is achieved that biases the final energy distribution in the network towards the source nodes. The relative rank algorithm in \cite{markov:white2003} and \cite{fogaras:personal2004} maintains many similarities to the particle propagation algorithm proposed in this article. At the end of the particle propagation algorithm, the relative energy between authors represents the relative competency of each author represented in the co-authorship network with respects to the manuscript. This section will first discuss an algorithm to construct a co-authorship network from a digital library repository and will then provide a formal representation of the particle-swarm algorithm used to locate referees in the resulting co-authorship network.

\subsection{Constructing a Co-Authorship Network}

A co-authorship network is defined by a graph composed of nodes that represent authors and edges that represent a joint publication between two authors \cite{newman:coauthor2004}.  Therefore, a co-authorship network is represented by the tuple $G=(N,E,W)$, where $N$ is the set of nodes, one for each author, in the network, $E$ is the set of edges relating the various authors, and $W$ is the set of weights representing the strength of tie between any two collaborating authors. In other words, any edge, $e_{l,j}$, connects two authors, $n_l$ and $n_j$, with a respective weight of $w_{l,j} \in \mathbb{R}^+$. The edge weight between any two authors is determined by Eq. \ref{eq:edgeweight}. 

	\begin{equation}
		\label{eq:edgeweight}
	 		w_{l,j} = w_{j,l} = \sum_{\forall m \in M \;\mathrm{by}\; l,j} \frac{1}{A(m)-1}
	 \end{equation}
	 
This equation represents two considerations. First, when the total number of authors for a manuscript, given by the function $A(m)$, is high, the resulting co-authorship weights will be low since the weight is distributed amongst the full of set of collaborating authors. This is represented by the fraction $\frac{1}{A(m)-1}$ where $A(m)$ returns the total number of authors for manuscript $m$. Second, the more frequently two authors co-author in the bibliographic record, the higher their co-authorship weight. The latter is represented by the summation, $\sum_{\forall m \in M \;\mathrm{by}\; l,j}$, where $M$ denotes the set of all manuscripts in a collection and $m \in M$. This method of co-authorship network construction is borrowed from \cite{coauth:liu2005,newman:science22001,newman:science2001}. The co-authorship network construction algorithm runs in $O(|M|)$.

The mentioned particle-swarm algorithm computed on the co-authorship network is a random process that requires the outgoing edge weights of a node to be represented as a probability distribution.  Therefore, the co-authorship edge weights must be normalized such that $\sum_{\forall e_{l,j} \in \mathrm{out}(n_l)} w_{l,j} = 1$ where $\mathrm{out}(n_l)$ is the set of outgoing edges from node $n_l$.

\subsection{Propagating a Particle-Swarm}

The purpose of the particle-swarm algorithm is to map a manuscript to a set of potential referees. Since a co-authorship network only expresses the relationship between authors, a manuscript will be represented as the set of authors in the manuscript's bibliography. Let the set $Q$ represent the set of authors cited in the bibliography of a particular article. For every author element $n_l \in Q$, there exists a corresponding unique node in the co-authorship network.  Therefore, $Q \subseteq N$. A distribution of particles, $P$, start their journey at $Q$ and propagate over the co-authorship network via the network edges. Any particle, $p_i \in P$, is composed of three components: an energy value, a energy decay property, a pointer to its current nodal location.

	\begin{enumerate}
		\item $\epsilon_i(t) \in \mathbb{R}$: is the amount of energy contained within the particle $p_i$ at time $t$
		\item $\delta_i \in [0,1]$: is the decay parameter governing the loss of energy as the particle $p_i$ propagates through the network
		\item $c_i(t) \in N$: is the location of the particle $p_i$ at time $t$
	\end{enumerate}

Every node in the co-authorship network has an accompanying energy value represented by a scalar within the energy vector $\bold{e} \in \mathbb{R}^{|N|}$. For instance, node $n_l$'s energy value is $\bold{e}_l$. The energy value for a node is incremented, or decremented, as particles traverse the node. At time $t=1$ there exists an energy distribution only over the set $Q$ such that for all $n_l \in Q$, $\bold{e}_l(1) > 0$. This means that at $t=1$, only those author nodes that are references in the manuscript contain an energy value greater than $0$. Furthermore, the more often a particular author is referenced by the manuscript, the more particles that author's node will initially receive at $t=1$. Therefore, if author $n_l$ is referenced once and author $n_j$ is referenced twice, then $n_j$ will have twice as many initial particles.

A particle moves through the co-authorship network by randomly selecting one outgoing edge from its current node, $c_i(t)$. The edge that is chosen is biased by the outgoing probability distribution where higher weighted edges have a higher probability of being chosen for traversal by the particle.  This function is represented as $\theta : \mathrm{out}(c_i(t)) \rightarrow e_{l,j}$.  At each time step a particle propagates to a neighboring node and updates the current node's energy value, $\bold{e}_{c_i(t)}$ according to Eq. \ref{eq:updateenergy}.

	\begin{equation}
		\label{eq:updateenergy}
	 		\bold{e}_{c_i(t)}(t+1) = \bold{e}_{c_i(t)}(t) + \epsilon_i(t)
	 \end{equation}

Once the particle has deposited its current energy value, it decays the energy value according to $\delta_i$ before moving to the next node in the network. This is represented by Eq. \ref{eq:particledeath}, where $k$ is a tunable parameter limiting the number of steps a particle is allowed to propagate.

	\begin{equation}
		\label{eq:particledeath}
			\epsilon_i(t+1) =
				\begin{cases}
					(1-\delta_i)\epsilon_i(t) & \mathrm{if} \; t \leq k\\
					0 & \mathrm{otherwise}
				\end{cases}
	 \end{equation}

such that at the final time step $k$\\

	\begin{equation}
		\label{eq:energyvectorsum}
			\bold{e}_l(k) = \sum_{t=1}^{t \leq k} \sum_{i=1}^{i \leq |P|}				
			\begin{cases}
					(1-\delta_i)^{t-1}\epsilon_i(1) & \mathrm{if} \; c_i(t-1) = n_l\\
					0 & \mathrm{otherwise}
			\end{cases}
	 \end{equation}

\begin{figure}[h!]
			\begin{center}
				\scalebox{0.5}[0.5]{
				\includegraphics[width=0.9\textwidth]{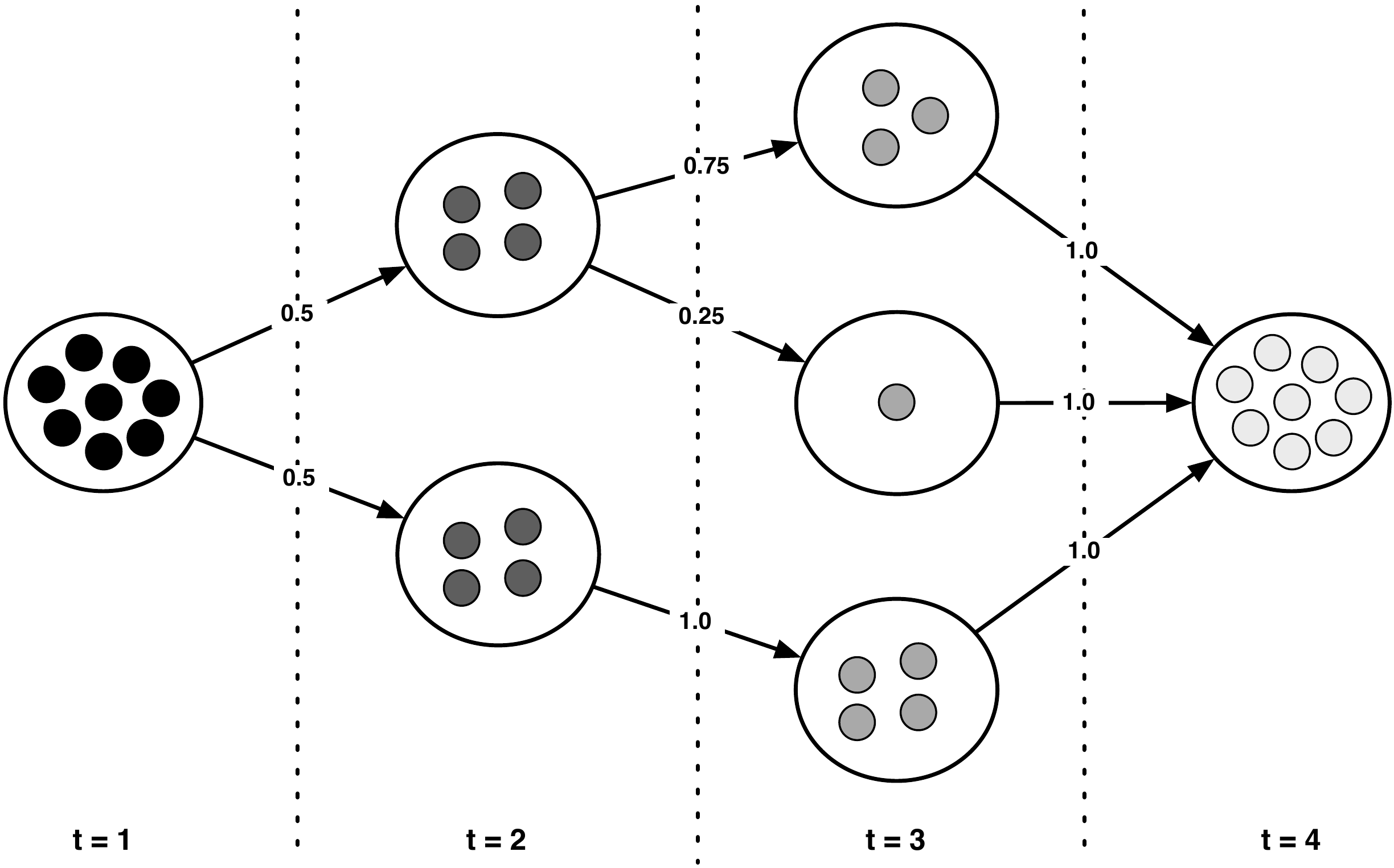}}
			\caption{\label{fig:peer-series} An example of decaying particles propagating in a probabilistic network}
			\end{center}
		\end{figure}

The running time of the particle propagation algorithm is $O(|P|k)$. Figure \ref{fig:peer-series} demonstrates how an initial distribution of particles propagates through a probabilistic network.  For each edge that a particle traverses, the local energy content, $\epsilon$, of each particle is decayed.  This is represented as the gray scale transition in the diagram. In Figure \ref{fig:peer-series}, the node at $t=4$ has less energy than the node at $t=1$ even though their respective particle populations are identical.

The particle-swarm algorithm propagates the initial $Q$ energy distribution over the co-authorship network such that at time $t=k$, for every node $n_l \in N$ that has a $\bold{e}_l(k) > 0$, $n_l$ is considered a potential referee for the manuscript. This set of potential referees is represented as the set $R = \{n_l \; | \; \bold{e}_l(k) > 0\}$, where $R \subseteq N$. Therefore, the particle-swarm algorithm maps a set of authors (references in the original manuscript $Q$) to a set of authors (referees in $R$) within the co-authorship network, $f : Q \rightarrow R$. A normalization of the energy vector, Eq. \ref{eq:normalizeenergy}, provides a membership value for each node in $R$ where $\mathrm{max}[\bold{e}(t)]$ returns the largest value in $\bold{e}$ and $\bold{e}_l(k+1) \in [0,1]$.

	\begin{equation}
		\label{eq:normalizeenergy}
	 		\bold{e}_l = \frac{\bold{e}_l(k)}{\mathrm{max}[\bold{e}(k)]}
	 \end{equation}

The pseudo-code for the particle-swarm algorithm is presented in Algorithm \ref{alg:particle}. With the initial particle distribution component the complete running time of the algorithm is $O(|P|+|P|k)$\footnote{In our test implementation, for a single article using the DBLP, the average run-time was 1.674 seconds on Intel Core Duo using Java 1.5.}.The particle-swarm algorithm, as used in this context, is a relative-rank algorithm \cite{markov:white2003}. The set of nodes in $N$ are ranked relative to $Q$. This is similar, though a more general case of finding the primary eigenvector of the network where the set of nodes in $N$ are ranked relative to $N$, $\delta = 0.0$, and $k \rightarrow \infty$.

	\incmargin{0.5cm}
	\restylealgo{boxed}
	\linesnumbered
	\begin{algorithm}[ht!]
	\begin{scriptsize}
	\Indp
			\CommentSty{\#distribute particles: $O(|P|)$}\;
			int $i = 1$\;
			\ForEach{$(n_l \in Q)$}{
				int $particlesPerNode = 100$\;
				\For{$($$j=0$, $j<particlesPerNode$, $j$++$)$}{
					$\epsilon_i = 1.0$; $\delta_i = 0.15$; $c_i = n_l$\;
					$i$++\;
				}
			}
			\BlankLine
			\CommentSty{\#propagate particles: $O(|P|k)$}\;
			int $t = 1$\;
			\While{$(t \leq k)$}{
				\For{$($$i=0$, $i < |P|$, $i$++$)$}{
					\If{$(\epsilon_i > 0)$}{
						$\bold{e}_{c_i} = \bold{e}_{c_i} + \epsilon_i$\;
						$\epsilon_i = (1 - \delta_i) \ast \epsilon_i$\;
			 			\If{$($$|\mathrm{out}(c_i)| == 0$$)$}{
							$\epsilon_i = 0$
							}
						\Else{
							$c_i = \theta(\mathrm{out}(c_i))$\;
						}
					}
				}
			$t$++\;
			}
		\caption{Particle-Swarm algorithm \label{alg:particle}}
	\end{scriptsize}
	\end{algorithm}
	\decmargin{0.5cm}

\subsection{The Particle-Swarm Parameter Space}

\begin{figure}[h!]
			\begin{center}
				\scalebox{0.5}[0.5]{
				\includegraphics[width=0.9\textwidth]{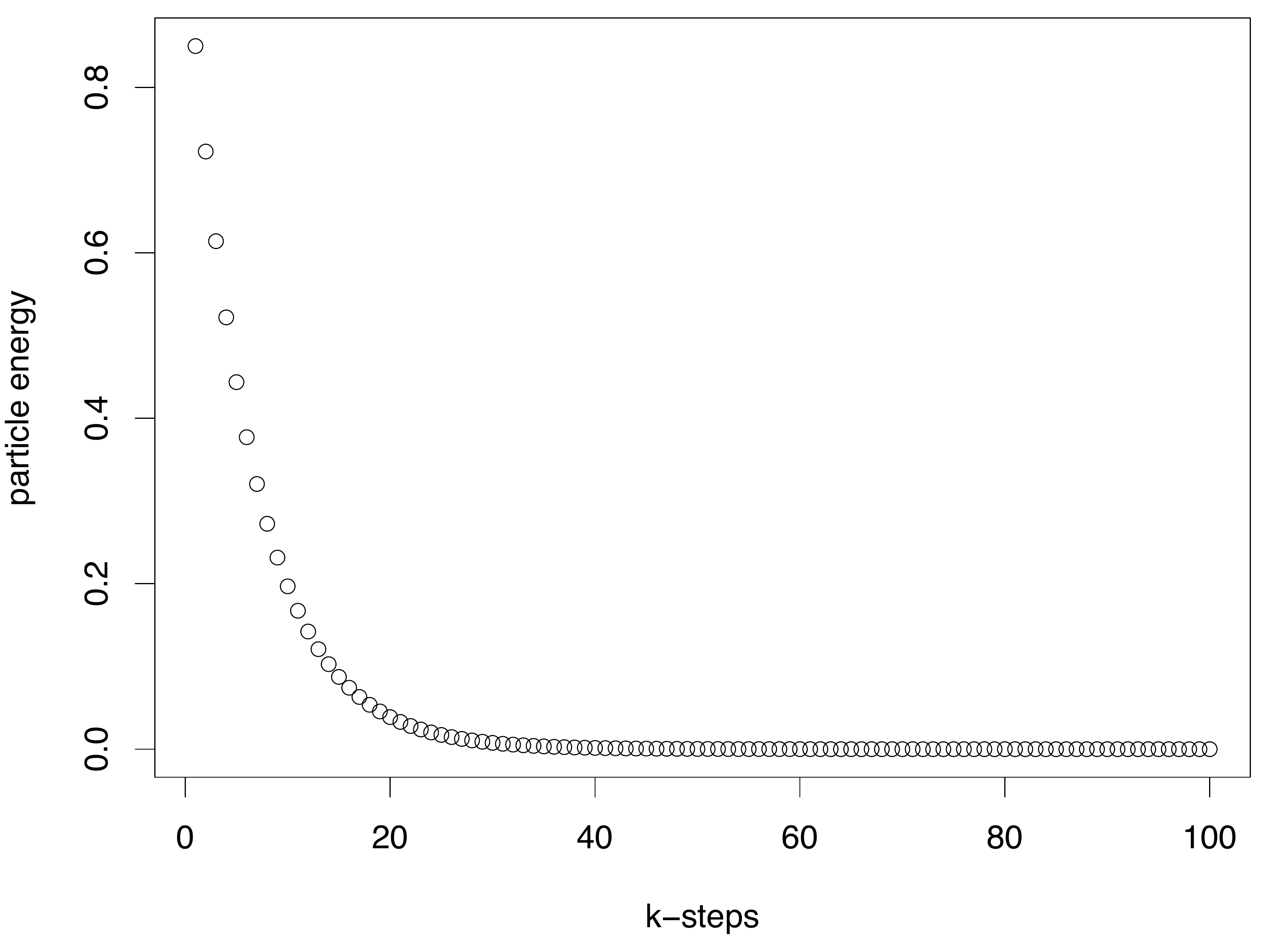}}
			\caption{\label{fig:energy-decay} Particle energy over $k$ for $\delta = 0.15$}
			\end{center}
		\end{figure}

There are three tunable parameters to the particle-swarm algorithm: the initial particle population $|P|$, the decay parameter $\delta$, and the number of steps for propagation $k$. The particle population can either be small in order to simulate a discrete random walker process or large to simulate a continuous spreading activation process. For the purpose of this study, we were more interested in the latter process. Furthermore, by increasing the initial particle population size, the random effects of the stochastic particle propagation algorithm are reduced. Our initial particle population for a single reference was $100$ particles. If an author is referenced more than once, then their initial particle population was $100x$ where $x$ is the number of references to that author. The parameter $k$ and $\delta$ have a similar effect on the network. If $\delta$ is high, then the amount of energy in the network as $k$ increases drops quickly since decay is a geometric progression with a negative common ratio. Thus, as $k \rightarrow \infty$, the effect of the particles on the final energy distribution diminishes to near $0$. For this reason, we set $k$ to $100$ since at $100$ steps, the amount of energy in a particle is $8.74 \times 10^{-8}$ and thus nearly equivalent to an infinite $k$. Energy over $k$ for $\delta=0.15$ is diagrammed in Figure \ref{fig:energy-decay}.

For the our experiments, we simply tuned $\delta$ and found an appropriate decay at $0.15$. However, when applying this algorithm to a different data set, various parameter space search algorithms can be used in association with human validation to find the most appropriate $\delta$ parameter for that particular community.

\section{Validating the Proposed Referee Identification Algorithm}

The $77$ members of the 2005 JCDL program committee are asked to indicate their reviewing preferences in advance of the reviewing assignments, i.e. they bid on the submissions they wish to review. While there were $281$ submissions to the 2005 JCDL, only $124$ submissions had bid data for all program committee members. When bidding, the PC members can choose from the following bid codes:

\begin{description}
	\item[1]I am an expert in the domain of the submission and want to review
	\item[2]I am an expert in the domain of the submission
	\item[3]I am not an expert in the domain of the submission
	\item[4]There exists a conflict of interest
\end{description}

The 2005 JCDL bid data provides a complete overview of which PC member actually volunteered to review which submissions. Ideally, the algorithm's referee predictions for a particular manuscript should correspond with the 2005 JCDL PC members that volunteered to review the same manuscript. Our evaluation of the effectiveness of the proposed referee identification algorithm therefore rests on a comparison of the particle energy values a PC member receives and their actual bid codes.

\begin{figure*}[ht!]
			\begin{center}
				\scalebox{1.0}[1.0]{
				\includegraphics[angle=0,width=0.45\textwidth]{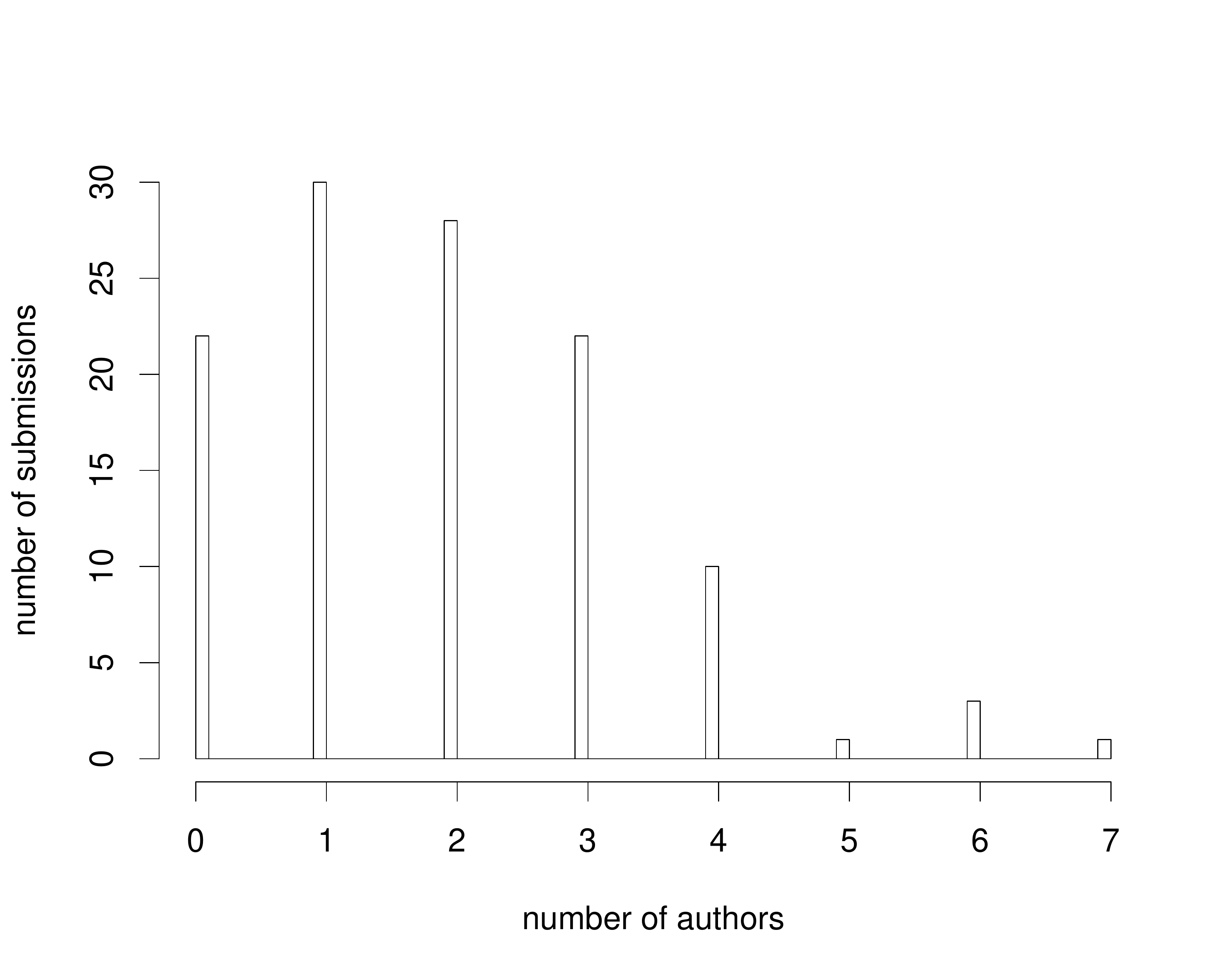}
				\includegraphics[angle=0,width=0.45\textwidth]{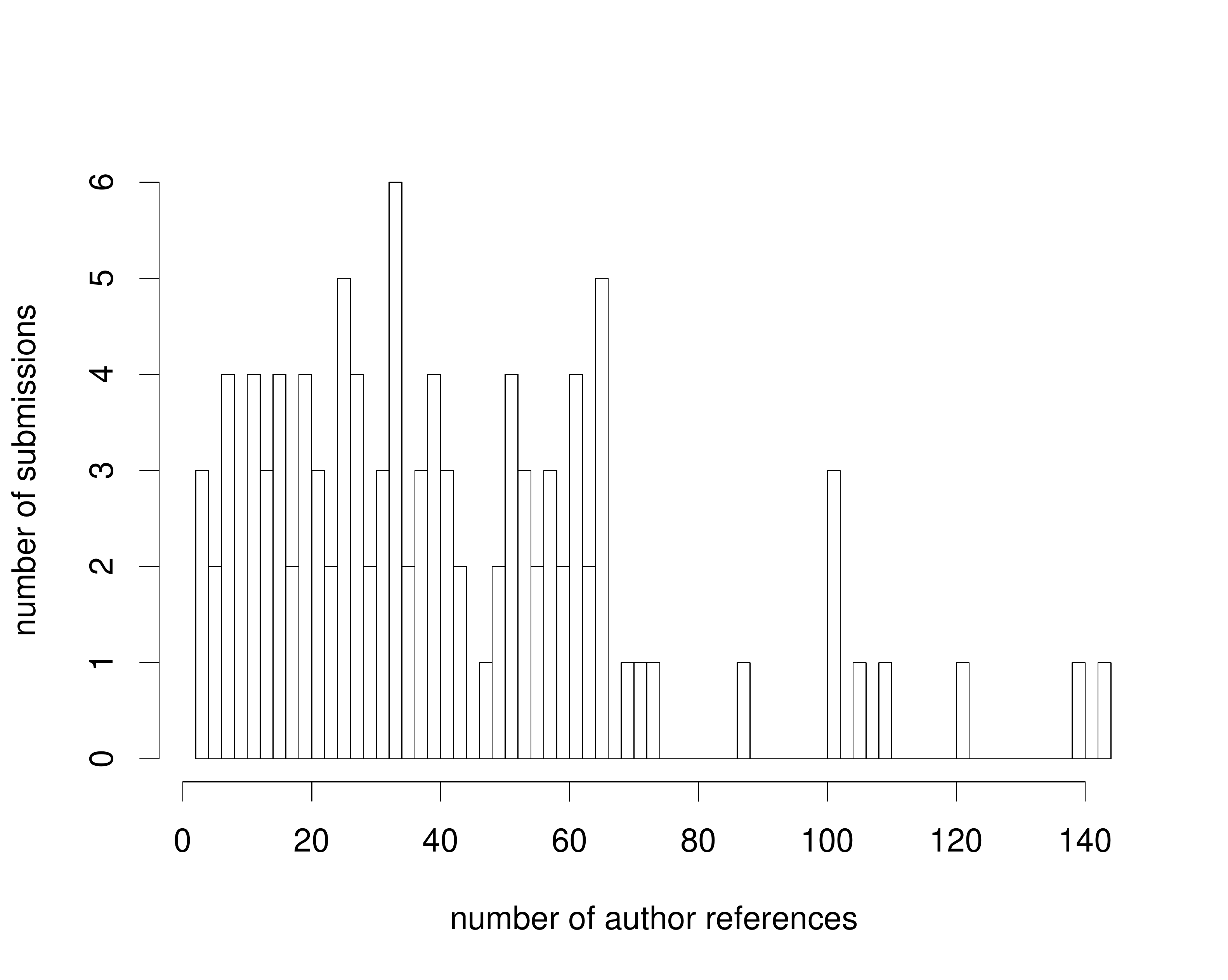}}
			\caption{\label{fig:data-dist} a.) authors per paper found in the DBLP b.) referenced authors per submission found in the DBLP}
			\end{center}
		\end{figure*}

The algorithm requires a co-authorship network to generate sets of potential referees. The co-authorship network chosen for this experiment was constructed using the Digital Bibliography and Library Project\footnote{DBLP available at: http://www.informatik.uni-trier.de/~ley/db/} (DBLP) bibliographic dataset.  This dataset is composed mainly of computer science journal and conference manuscripts (for which the digital library agenda is a sub-domain). The constructed network has 284,082 author nodes and 2,167,018 co-authorship edges. Of the $77$ PC members, $8$ were not found within the DBLP. Thus, 89\% of the PC members were found in the DBLP. For those members not in the DBLP, their bid behavior was excluded from the following analysis. Furthermore, $22$ articles did not have identifiable authors in the DBLP. Thus, only 83\% of the articles with bid data had authors in the DBLP. Figure \ref{fig:data-dist} diagrams the distribution of authors and author references found in the DBLP. Finally, no advanced name disambiguation algorithm was used. Only when the last name, first initial, and middle initial match did we consider that a positive identification.

This section will first discuss the general methodology of the algorithm validation and then provide the results of a comparison of the 2005 JCDL bid codes and the algorithms referee predictions for the 2005 JCDL submissions.

\subsection{Methodological Overview}

The proposed referee identification algorithm can be said to produce valid results if its referee predictions match the actual 2005 JCDL PC bid codes. For example, a PC member who entered bid code 1 (expert wanting to review) for a particular manuscript should ideally receive a higher particle energy value than a PC member who entered bid code 3 (not an expert). Since this should be the case for all manuscripts, the overall effectiveness of the algorithm can be determined by summing the energy values of all PC members who entered a particular bid code and comparing the resulting total energy values across bid codes. This means that PC members whose bids indicate they are experts (bid codes 1 and 2) should receive significantly higher energy values over all submissions than those whose bids indicated they are not experts (3). If this is the case, it can be said the algorithm's particle energy values successfully predict which PC member should be refereeing a particular manuscript.

In fact, if we'd denote the total particle energy $e$ assigned to any particular bid code $b$ as $e_b$, then the final distribution of particle energy most indicative of the effectiveness of the referee identification and weighting algorithm would be

\begin{equation}
	e_1 \approx e_2 > e_3 \approx e_4.
\label{bidenergy}
\end{equation}

\begin{figure}[ht!]
			\begin{center}
				\scalebox{0.5}[0.5]{
				\includegraphics[width=0.9\textwidth]{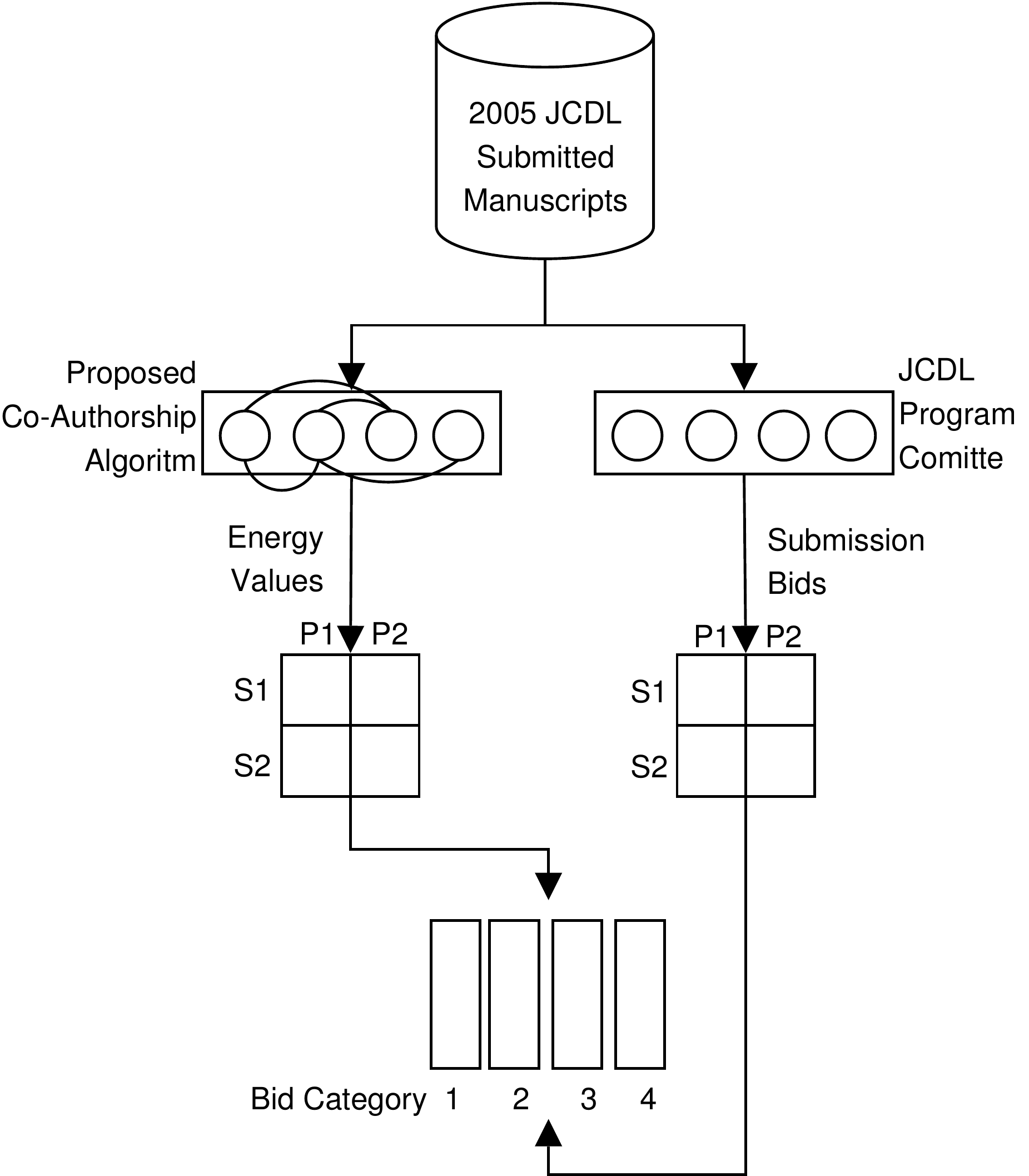}}
			\caption{\label{fig:methodology} Methodology for validating the proposed algorithm}
			\end{center}
		\end{figure}

The idea of matching particle energy assignment to actual PC member bid codes is outlined in Figure \ref{fig:methodology} where S1 refers to submission number 1 and P1 refers to program committee member number 1.

To test the degree to which PC member bid codes and the proposed algorithm's particle energy values overlap, each submitted manuscript in 2005 JCDL submission archive is parsed to extract its references using the Paracite\footnote{ParaCite available at: http://paracite.eprints.org/developers/} toolkit.  The referenced authors in the DBLP co-authorship network are then each supplied with $100$ particles where $\epsilon=1.0$, $\delta=0.15$, and $k=100$. At $k=100$, the energy level of a particle is near zero, $(1-0.85)^{100}$. The particle-swarm algorithm propagates the initial positive energy from the submission's bibliographic reference nodes to other scientists in the DBLP co-authorship network via the network edges as described in the previous section (Algorithm \ref{alg:particle}). The generated particle energy  for each PC member is recorded and added to the particular PC member's bid code for that manuscript. The accumulated particular energy values for each bid code can then be examined to determine how well they match the inequality given by Eq. \ref{bidenergy}.

\subsection{The Results of the Proposed Algorithms}

\begin{figure*}
			\begin{center}
				\scalebox{0.9}[0.9]{
				\includegraphics[angle=0,width=0.45\textwidth]{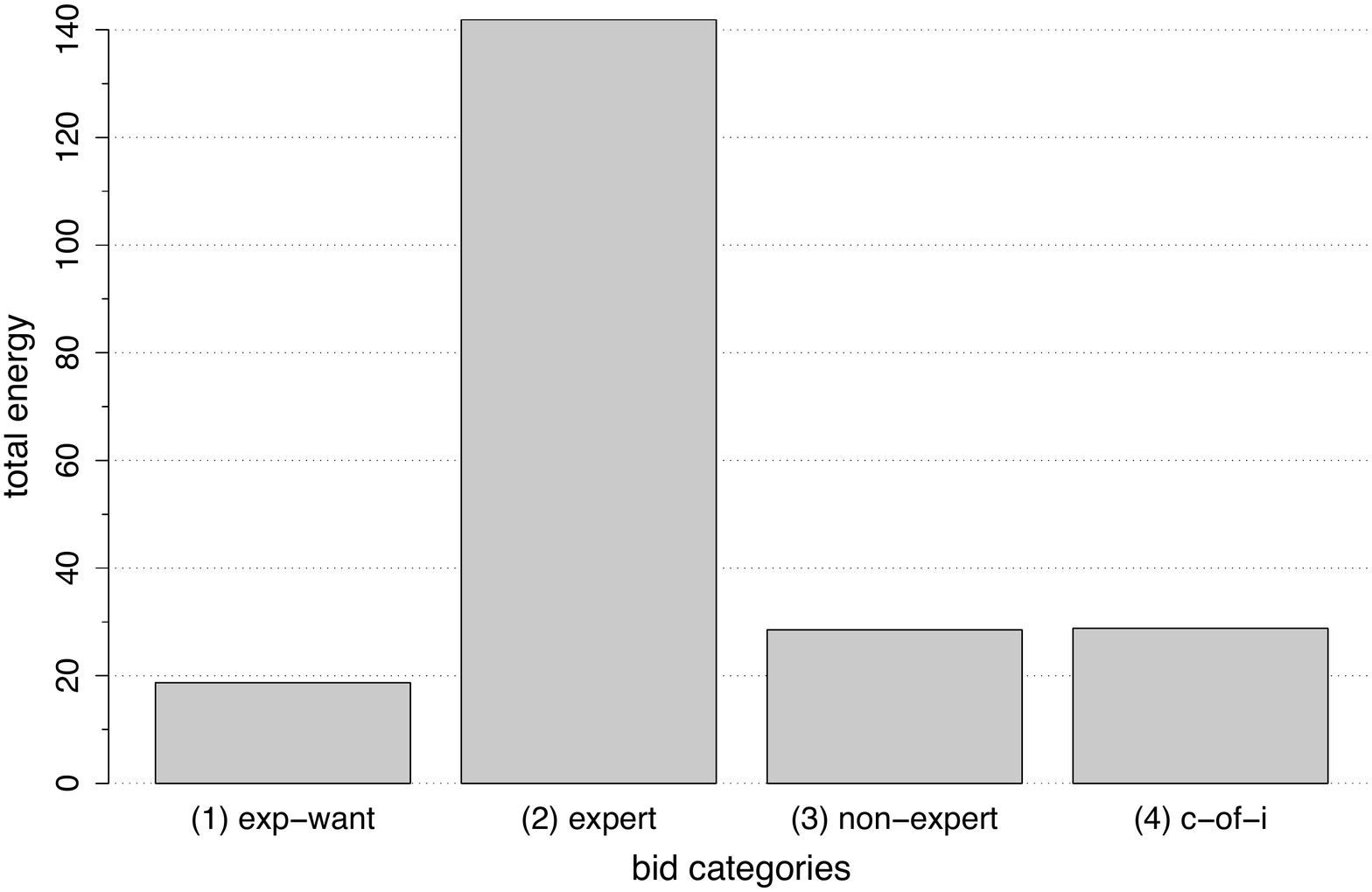}
				\includegraphics[angle=0,width=0.45\textwidth]{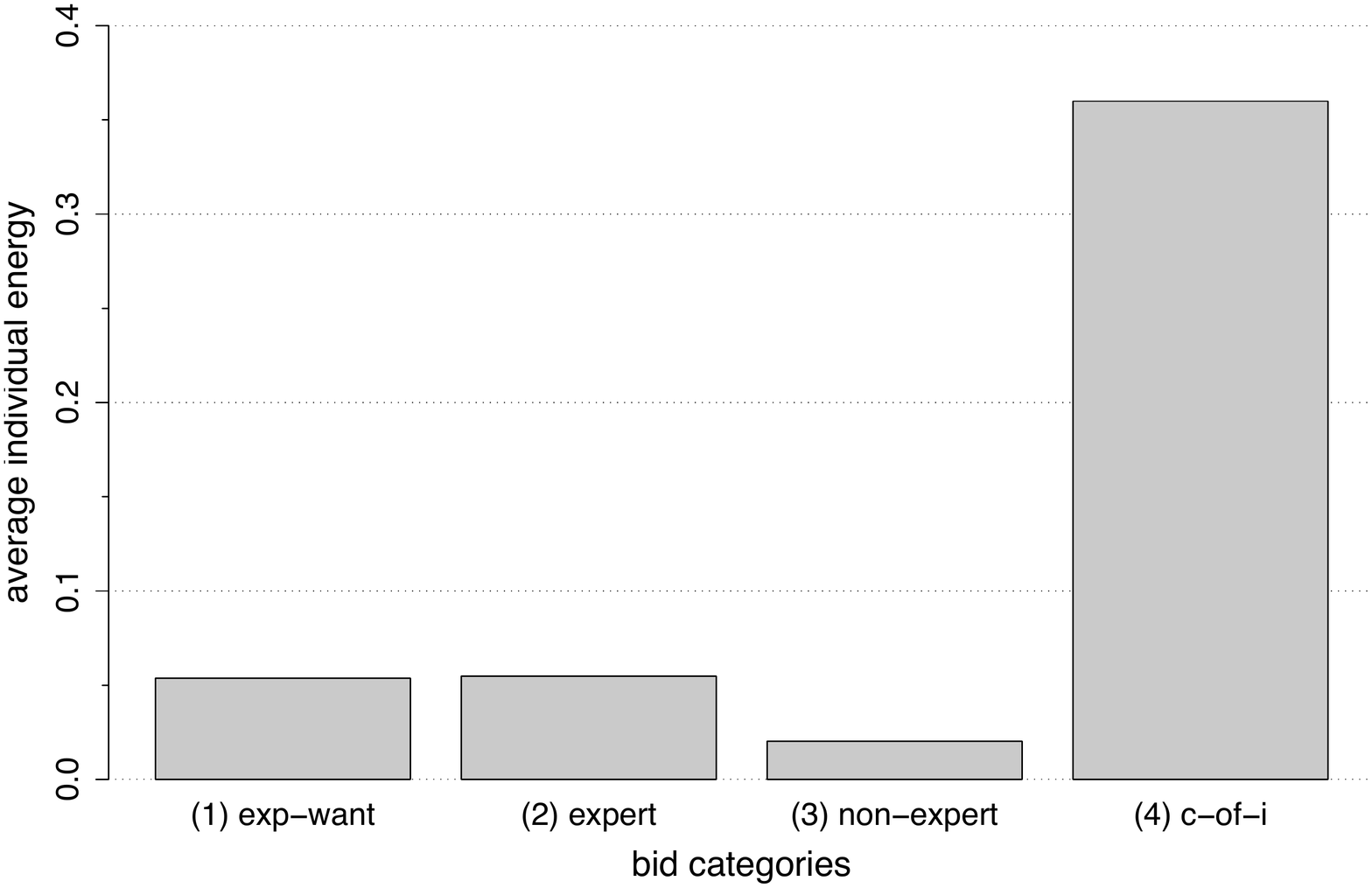}}
			\caption{\label{fig:coauthor-sumk0} Total energy in the various bid categories and mean energy in the various bid categories.}
			\end{center}
		\end{figure*}

\begin{figure*}
			\begin{center}
				\scalebox{1}[1]{
				\includegraphics[angle=0,width=0.8\textwidth]{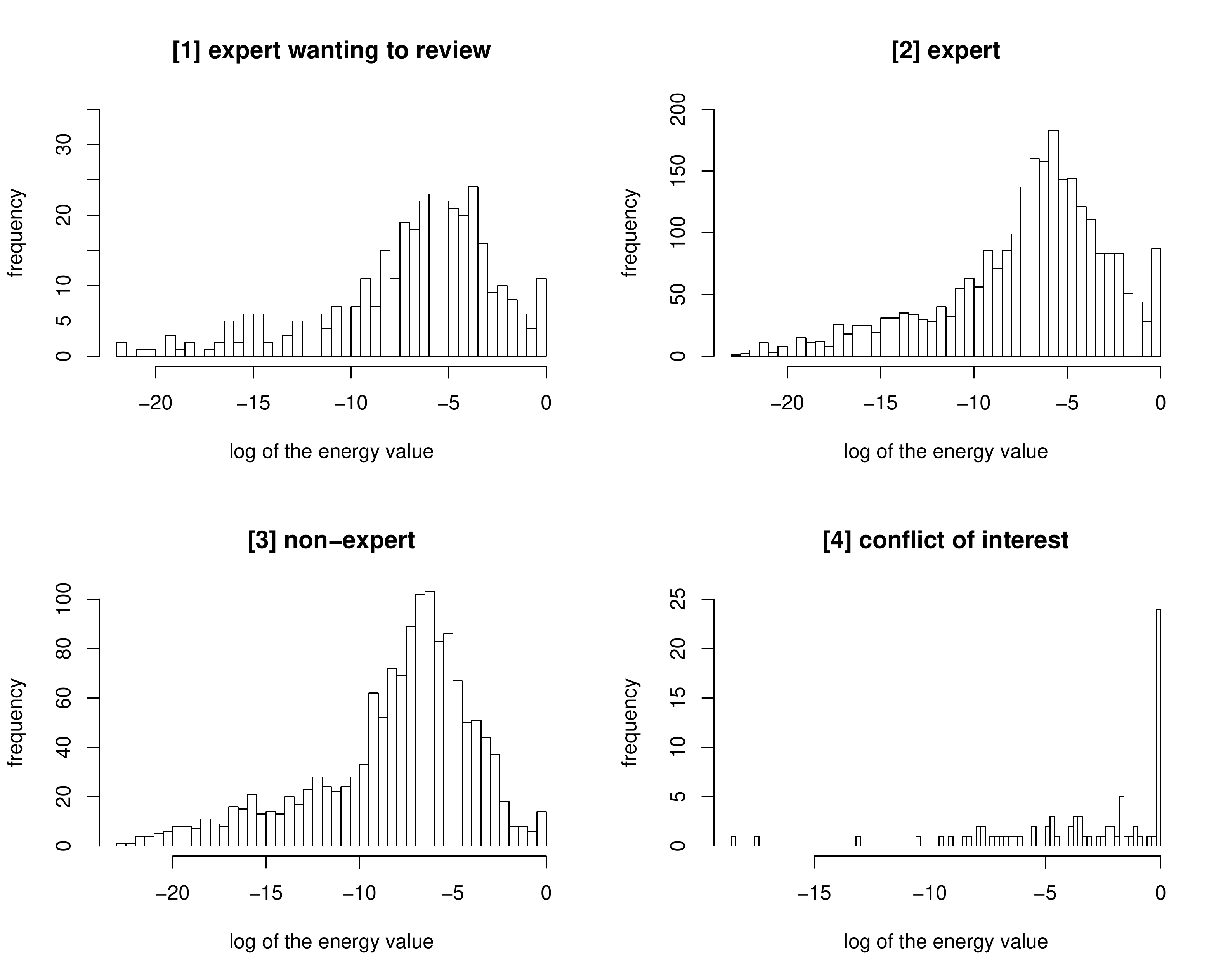}}
			\caption{\label{fig:coauthor-distk0} Distribution of energy in the various bid categories in a log-normal plot}
			\end{center}
		\end{figure*}

Particle energy values were generated for the entire 2005 JCDL submission archive and compared to the PC members bid codes. Figure \ref{fig:coauthor-sumk0} provides the total amount of energy each referee bid group received over all $124$ submissions as well as the mean energy for each bid category. Figure \ref{fig:coauthor-distk0} plots the frequency of the various energy values in the different bid groups. The $x$-axis of Figure \ref{fig:coauthor-distk0} represents a range of energy values and the $y$-axis represents the number of PC members in that bid group that fall within a particular range of energy.
		
		\begin{table}[hptb]
		\begin{center}
		\begin{scriptsize}
			\begin{tabular}{|c||c|c|c|c|}\hline
				\textbf{bid}&\textbf{1}&\textbf{2}&\textbf{3}&\textbf{4}\\\hline\hline
				\textbf{1}&1.0     &0.211    &$<0.001$ & $<0.001$ \\\hline
				\textbf{2}&0.211    &1.0     &$<0.001$&$<0.001$\\\hline 
				\textbf{3}& $<0.001$  &$<0.001$&1.0     &$<0.001$\\\hline
				\textbf{4}& $<0.001$&$<0.001$&$<0.001$&1.0\\\hline
			\end{tabular}
		\caption{\label{tab:pvalues}Kolmogorov-Smirnov p-values for each bid category pairs}
		\end{scriptsize}
		\end{center}
		\end{table}

A Kolmogorov-Smirnov non-parametric test between the energy values of the different bid categories was performed \cite{nonpar:conover1971}. Table \ref{tab:pvalues} provides the p-values. In line with the hypothesis, the proposed referee identification algorithm is able to make a statistically significant distinction between expert, non-expert, and conflict of interest referees. The algorithm, however, cannot make a significant distinction between experts and experts wanting review (bid groups $1$ and $2$).  This could mean that the co-authorship network does not contain information about current research interest of a scientist, only their domain of expertise.

The results demonstrate that conflict of interest referees are assigned a significant amount of energy. This would be expected since conflict of interests are usually closely related in expertise to the author of the submission (i.e.~are the author themselves or have co-authored with the author previously). The reason that authors of the submission receive an excessive amount of energy is due in large part to the fact that authors cite themselves more often than not and therefore would receive a high energy amount with respect to their own manuscript. Individuals who have co-authored with the authors of the submission (those individuals one step away from the authors in the co-authorship network) would also tend to receive a large amount of energy. If energy is a measure of the amount of decision-making influence that a referee should have with respects to the manuscript then it is desirable to ensure that conflict of interest referees receive no positive particle energy. Therefore, the next section will provide a modification to the proposed algorithm in order to reduce the amount of energy that conflict of interest referees receive.

\subsection{Conflict of Interest Reduction by Negative Particle Energy}

\begin{figure}[ht!]
			\begin{center}
				\scalebox{0.5}[0.5]{
				\includegraphics[width=0.8\textwidth]{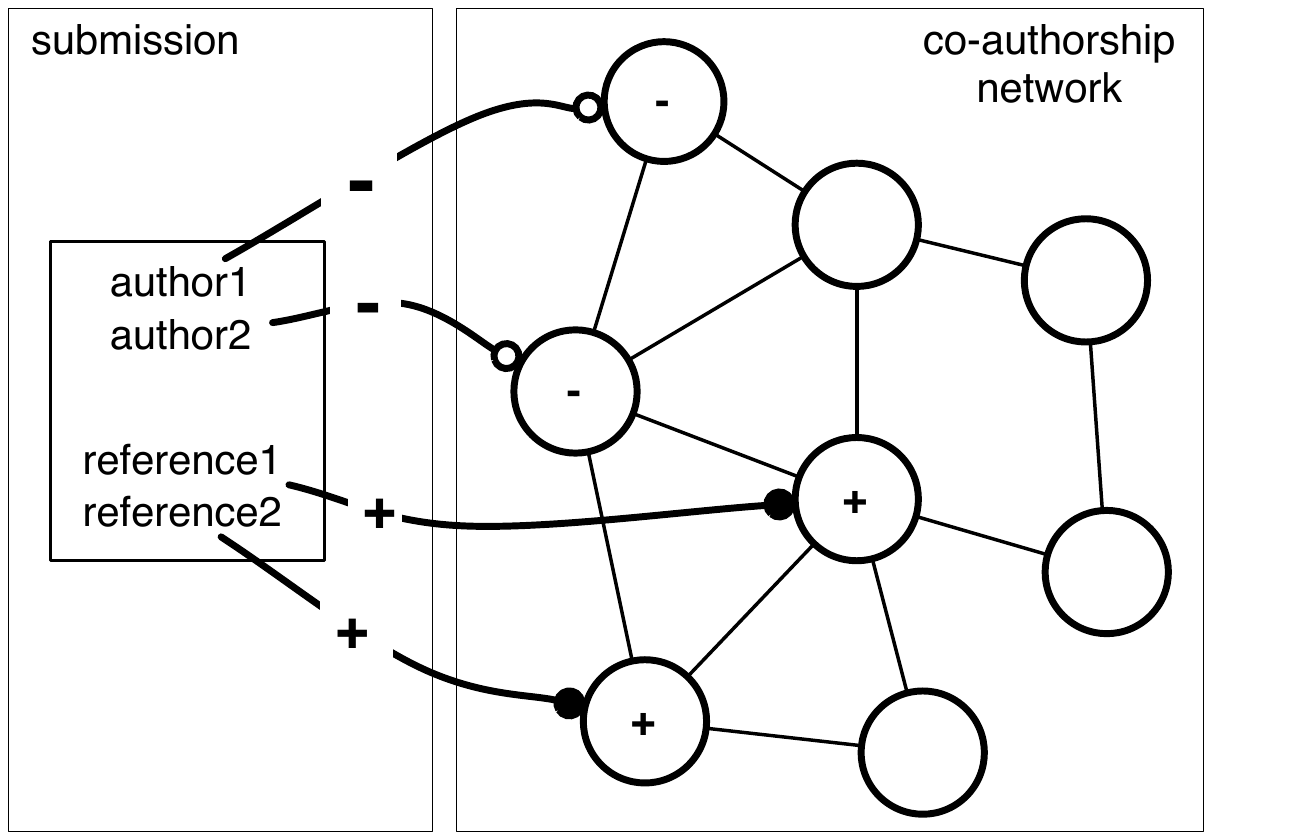}}
			\caption{\label{fig:peer-swarm} The application of positive and negative energy particle-swarms}
			\end{center}
		\end{figure}

This section outlines an extension to the algorithm aimed at reducing the degree to which conflict of interest referees receive particle energy. In the modified algorithm, a negative energy swarm is placed at the submission author nodes as shown in Figure \ref{fig:peer-swarm}.  This negative energy particle-swarm will negate the energy otherwise assigned to the manuscript authors themselves and those individuals most closely related. It is hypothesized that this will reduce the amount of energy received by conflict of interest referees.

A negative energy particle was defined with the following properties: $\epsilon=-1000.0$, $\delta=0.0$. Obviously, if the co-authorship network is connected, then a `black-out' swarm with no decay that can propagate indefinitely will remove all positive energy in the network. Therefore, the propagation depth or steps, $k$, of the negative energy particles is varied to control the neighborhood in which their inhibitive effects take place.

\begin{figure*}[ht!]
			\begin{center}
				\scalebox{0.9}[0.9]{
				\includegraphics[angle=0,width=0.5\textwidth]{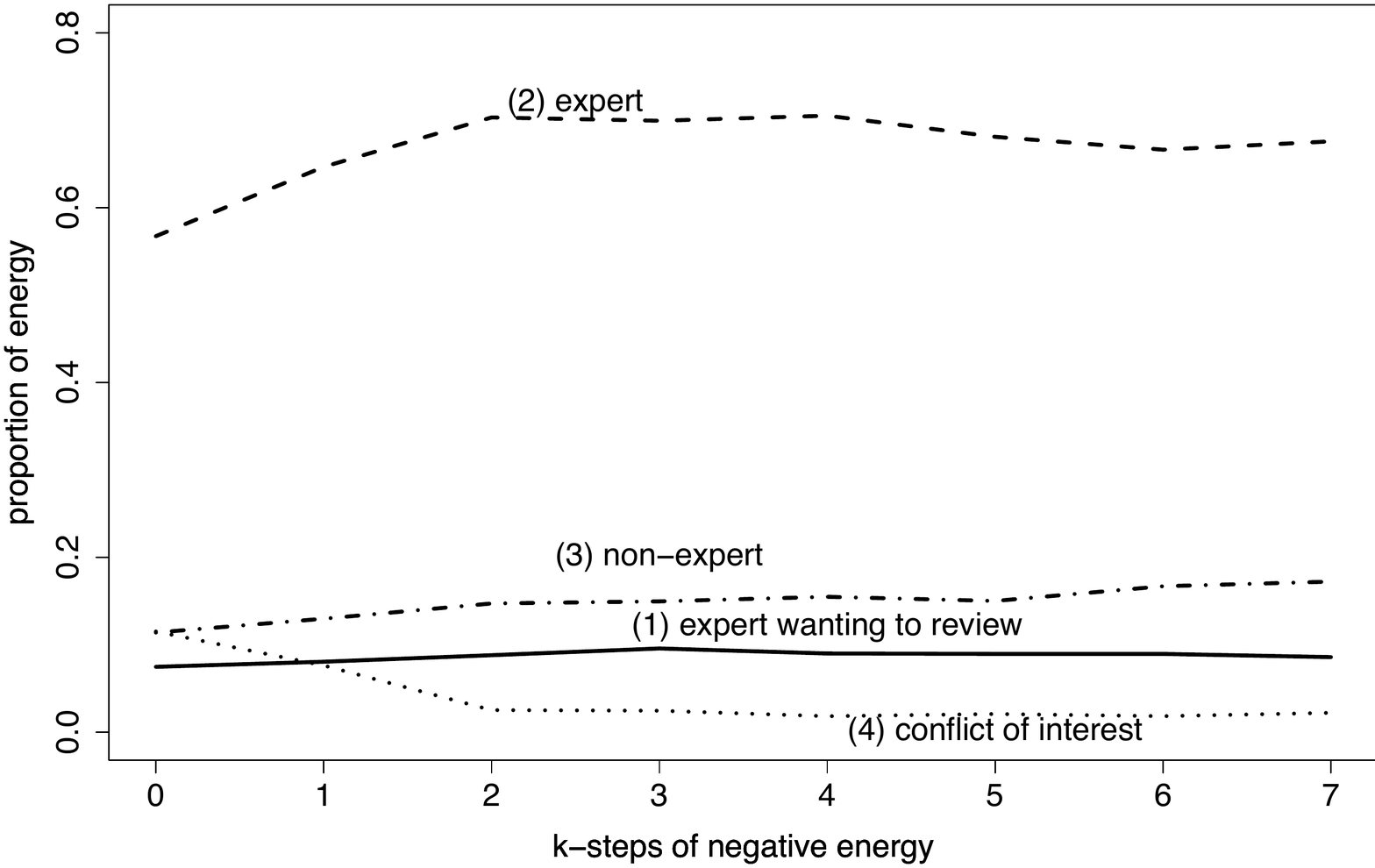}
				\includegraphics[angle=0,width=0.5\textwidth]{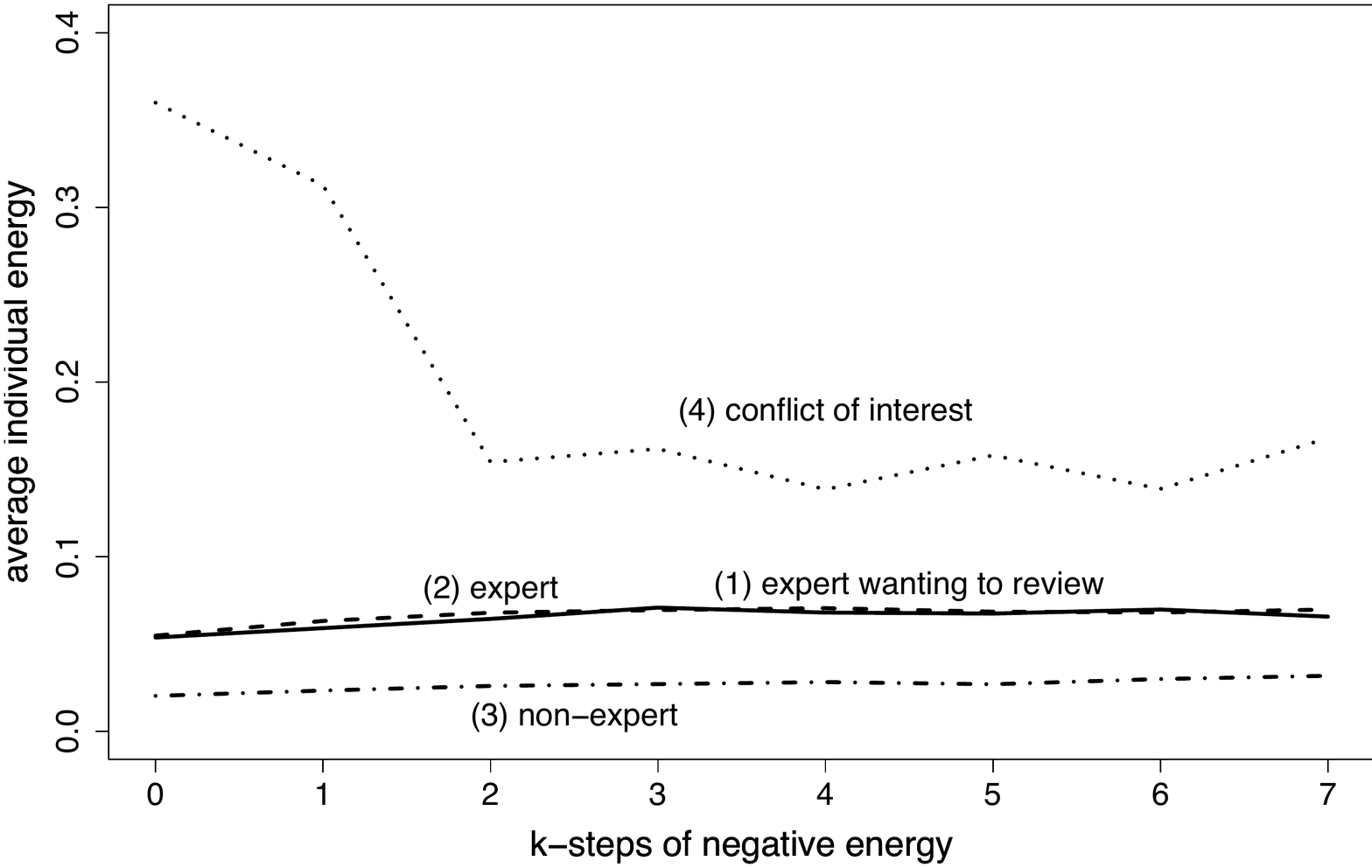}}
			\caption{\label{fig:coauthor-sum} A `black-out' distribution for varying $k$ and the mean distribution over the bid categories.}
			\end{center}
		\end{figure*}

Figure \ref{fig:coauthor-sum} denotes the total amount of energy for all submissions in each bid category after $k$ number of `black-out' propagations and the average energy for any one individual in that bid category. The more steps the swarm is allowed to propagate, the more energy removed from the network. Thus, it is important to stop that `black-out' swarm from removing all energy in the network. As presented in Figure \ref{fig:coauthor-sum}, the most optimal $k$, i.e. depth of propagation, for the negative energy particle-swarm is approximately $2$.  Indeed, at $k\approx2$, the proportion of energy located at expert referees is the greatest, and the proportion of energy located at conflict of interest and non-expert referees is the lowest. Note that when the propagation algorithm is complete, any node with less than $0$ energy has $0$ energy added to their respective bid category. It should be noted that the negative energy particles have the same effect on $\bold{e}$ as setting all nodes energy in the $k$-neighborhood of the author node(s) to $0$. However, in theory, since this is a stochastic process, it is possible for the `black-out' swarm to not reach all $k$ neighbors. Furthermore, $k=0$ is when no `black-out' is distributed to the manuscript's author node(s) and therefore is equivalent to the original version of the algorithm.

\begin{figure*}[ht!]
			\begin{center}
				\scalebox{1}[1]{
				\includegraphics[angle=0,width=0.8\textwidth]{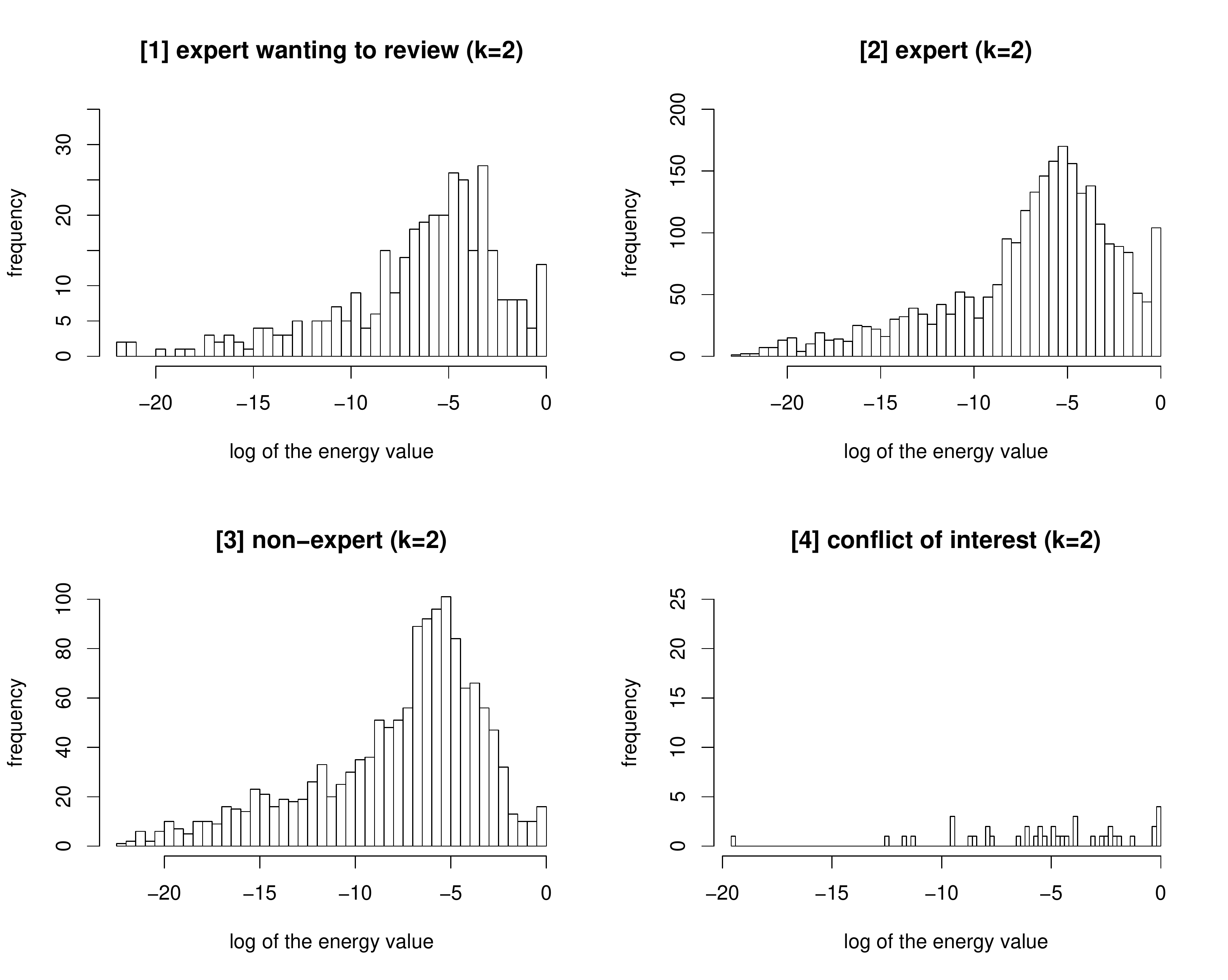}}
			\caption{\label{fig:group-hist} $k=2$ `black out' swarm energy distributions on log/linear plot}
			\end{center}
	\end{figure*}

Figure \ref{fig:group-hist} shows the energy distributions on a log/linear scale for the most optimal $k$ for the `black out' swarm. What is apparent is that for all referee types, except conflict of interest referees, the energy distribution remains relatively unchanged.  This further demonstrates that most conflict of interest referees are located, in the co-authorship network, in the vicinity of the submission's author(s) because as particle energy decays over time, the highest energy values are distributed early in the diffusion process. Table \ref{tab:pvalues2} present the p-values for the Kolmogorov-Smirnov of these energy distributions.
	
		\begin{table}[hptb]
		\begin{center}
		\begin{scriptsize}
			\begin{tabular}{|c||c|c|c|c|}\hline
				\textbf{bid}&\textbf{1}&\textbf{2}&\textbf{3}&\textbf{4}\\\hline\hline
				\textbf{1}&1.0     &0.3486    &$<0.001$ & 0.2187 \\\hline
				\textbf{2}&0.3486  &1.0     &$<0.001$&0.1795\\\hline 
				\textbf{3}& $<0.001$  &$<0.001$&1.0     &0.007\\\hline
				\textbf{4}&0.2187&0.1795&0.0072&1.0\\\hline
			\end{tabular}
		\caption{\label{tab:pvalues2}Kolmogorov-Smirnov p-values for each bid category pairs}
		\end{scriptsize}
		\end{center}
		\end{table}

Table \ref{tab:recall} presents the percentage recall of the bid members with greater than $0.0$ energy. As can be determined from the table, the `black-out' swarm is able to reduce the number of conflict of interest referees that are provided energy.

		\begin{table}[hptb]
		\begin{center}
		\begin{scriptsize}
			\begin{tabular}{|c||c|c|c|c|}\hline
				\textbf{bid/step}&\textbf{1}&\textbf{2}&\textbf{3}&\textbf{4}\\\hline\hline
				\textbf{0-step}& 0.734     &0.727    & 0.691 & 0.899 \\\hline
				\textbf{2-step}& 0.722  & 0.727    & 0.690 & 0.461 \\\hline 
			\end{tabular}
		\caption{\label{tab:recall}The percentage of recall of program committee members from the respective bid categories}
		\end{scriptsize}
		\end{center}
		\end{table}

Finally, in order to determine the highest energy referees for both the non- and `black-out' swarm, the top energy referee values were considered. Those referees that had a maximum energy of $1.0$ as identified by Equation \ref{eq:normalizeenergy} were removed. The number of $1.0$ energy referees is apparent from the respective Figures \ref{fig:coauthor-distk0} and \ref{fig:group-hist}.  Each bid category has a collection of $1.0$ referees as identified by right most bar in each plot of Figure \ref{fig:coauthor-distk0} and Figure \ref{fig:group-hist}. For all those with less than $1.0$ energy, the top $5$ energy values of each bid category is presented in Table \ref{tab:0steptop} for a $0$-step `black-out' and in Table \ref{tab:2steptop} for a $2$-step `black-out' swarm. Note that for journal situations where only 3 or 4 referees is desirable, the top 4 highest energy  referees are in bid category number 2 and 1 (i.e.~experts and experts wanting to review). 

		\begin{table}[hptb]
		\begin{center}
		\begin{scriptsize}
			\begin{tabular}{|c||c|c|c|c|c|}\hline
				\textbf{rank/bid}&\textbf{rank1}&\textbf{rank2}&\textbf{rank3}&\textbf{rank4}&\textbf{rank5}\\\hline\hline
				\textbf{1}& 0.958 & 0.933 & 0.928 & 0.851 & 0.765 \\\hline
				\textbf{2}& 0.996 & 0.987 & 0.982 & 0.976 & 0.948 \\\hline
				\textbf{3}& 0.978 & 0.941 & 0.906 & 0.872 & 0.793  \\\hline
				\textbf{4}& 0.942 & 0.705 & 0.617 & 0.409 & 0.335 \\\hline
			\end{tabular}
		\caption{\label{tab:0steptop}The energy values of the program committee members in their respective bid categories without the `black-out'.}
		\end{scriptsize}
		\end{center}
		\end{table}

		\begin{table}[hptb]
		\begin{center}
		\begin{scriptsize}
			\begin{tabular}{|c||c|c|c|c|c|}\hline
				\textbf{rank/bid}&\textbf{rank1}&\textbf{rank2}&\textbf{rank3}&\textbf{rank4}&\textbf{rank5}\\\hline\hline
				\textbf{1}& 0.974 & 0.948 & 0.926 & 0.920 & 0.843 \\\hline
				\textbf{2}& 0.980 & 0.965 & 0.965 & 0.953 & 0.952 \\\hline
				\textbf{3}& 0.862 & 0.848 & 0.848 & 0.780 & 0.778 \\\hline
				\textbf{4}& 0.872 & 0.729 & 0.671 & 0.252 & 0.155 \\\hline
			\end{tabular}
		\caption{\label{tab:2steptop}The energy values of the program committee members in their respective bid categories with `black-out' swarm of $k=2$.}
		\end{scriptsize}
		\end{center}
		\end{table}

\section{Future Research}

It can be concluded from Figures \ref{fig:coauthor-sum} and \ref{fig:group-hist}, that the `black-out' particle-swarm is able to remove a significant amount of energy from the conflict of interest referees. Unfortunately, not all conflict of interest referee energy is reduced to zero. This may be because co-authorship relationships are not the only reason that conflict of interest situations emerge. We can only speculate that the incorporation of other relational information such as affiliation data, funding networks and institutional networks might provide the necessary network edges that will allow the `black-out' particle-swarm to remove more of the conflict of interest referees. One could also conceive of a situation in which the algorithm generates a set of potential referees which are then vetted by human operators on the basis of extraneous information to identify and exclude conflict of interest referees. In spite of its propensity to identify conflict of interest referees, such an application would nevertheless greatly improve the referee identification process. This idea will be left to future research in this area.

It is important to further emphasize that this algorithm has only been validated on a co-authorship network that is focused on the computer sciences for which the digital library research agenda is a particular sub-domain. Different scientific disciplines will have different network topologies \cite{newman:coauthor2004} and therefore may require different particle-swarm parameters. Therefore, conflict of interest situations may not be so easily defined as those individuals 1 or 2 steps away in the co-authorship network.  We recommend that this algorithm, before being implemented within a specific community other than the digital library community, be validated using the methodology described in this paper.

The Digital Library Research and Prototyping Team at the Los Alamos National Laboratory is currently engineering the a massive semantic scholarly network \cite{semever:bollen2007}. This network will include relationships between authors, papers, journals, conferences, publishers, and institutions represented in a multi-billion triple RDF triple store. Future work in the area will allow us identify which relationships are most important in not only making this algorithm more accurate at identifying referees, but also conflict of interest situations. For one, various parameters of the algorithm will be tested to determine the role of prolificness of an author and how they effect the particle-swarm energy distribution. As authors write more papers, their connectivity and thus, the probability of being encountered by a particle increases. It may be important to understand how to adjust the algorithm to account for such aspects of a reviewer. The network model of the scholarly community will also include temporal information and thus, referee research trends could be taken into account to provide a mechanism of distinguishing between those referees in bid category $1$ and bid category $2$. Furthermore, the semantic network substrate will allow us to test various `semantically-aware' algorithms. For instance, the grammar-based particle-swarm algorithm  \cite{grammar:rodriguez2007} can be used to direct the particles along a semantically meaningful path and thus will provide us with a wide-range of metrics for which to compare and contrast. We will be able to survey the full landscape of network analysis algorithm such that we may identify which algorithms and which semantics provide the best mechanism for identifying peer-reviewers.

\section{Conclusion}

The peer-review process, in its present form, is mainly mediated by human efforts, i.e. authors, referees, and journal editors or conference organizers interact to produce a set of vetted, certified publications. This paper outlines an automatic referee identification algorithm that requires no human intervention, is computationally efficient, and can, to some extent, automatically identify conflict of interest situations. The referee weighting aspect of the algorithm provides a strong incentive for its use in open commentary peer-review. The level of automation provides the necessary infrastructure to decouple the publication process from the peer-review process in the sense that editors are no longer required to assign referees.  A system that uses such an algorithm to identify and weight its reviewers is more efficient as well as more equitable and objective while at the same time potentially allowing any member of the community contribute a review to a manuscript. Furthermore, a quantified peer-review service opens the peer-review process as an object of scientific inquiry.

We identify an inherent paradox associated with referee identification. On the one hand, it is important to locate the most qualified referees to review a manuscript, while on the other, it is important to remove conflict of interest referees from the review process. The paradox lies in the fact that many of the most qualified referees are necessarily conflict of interest referees. Therefore, an automated referee identification algorithm must achieve a balance between accepting qualified referees while at the same time rejecting conflict of interest referees. It can only be concluded that the current `honor system' will continue to play an important role in the peer-review process as no computer algorithm to date can accurately identify the social and political elements of conflict of interest situations of peer-review.

\section{Acknowledgments}

This research could not have been conducted if it were not for the support of the 2005 JCDL program chair and steering committee. Herbert Van de Sompel supported this research through data acquisition. Finally, we would like to thank the Journal of Memetics\footnote{Journal of Memetics available at: http://www.jom-emit.org/} for using a prototype implementation of the algorithm in their peer-review process. This research was financially supported by the Los Alamos National Laboratory.

\end{document}